# Homogenous $In_xGa_{1-x}N$ alloys on ZnO substrates: A new approach for high performance thermoelectric materials

Yining Feng[1,2], Evan Witkoske[3], Bahadir Kucukgok[1,2], Yee Rui Koh[3,4,5], Ali Shakouri[3,4], Ian T. Ferguson[6], and Na Lu[1,4,7*]

[1]*Lyles School of Civil Engineering, Purdue University, West Lafayette, IN 47907, USA;* luna@purdue.edu*

[2]*Applied Materials Division, Argonne National Laboratory, Lemont, IL 60439, USA*

[3]*School of Electrical and Computer Engineering, Purdue University, West Lafayette, IN 47907, USA*

[4]*Birck Nanotechnology Center, Purdue University, West Lafayette, IN 47907, USA*

[5]*Department of Mechanical and Aerospace Engineering, University of Virginia, Charlottesville, VA 22904, USA*

[6]*Electrical and Computer Engineering, Missouri University of Science and Technology, Rolla MO 65409, USA*

[7]*School of Materials Engineering, Purdue University, West Lafayette, IN 47907, USA.*

## Abstract

High performance thermoelectric materials for wide-range temperature applications still remains a challenge. In this study, we have produced high-quality homogeneous $In_{0.32}Ga_{0.68}N$ on ZnO substrates, with no phase separation at high Indium content, using metal organic chemical vapor deposition for thermoelectric applications. A record high room temperature figure of merit zT is obtained of 0.86, which is five times larger than that of SiGe, the current state of the art high temperature thermoelectric material. These materials are shown to have a nearly perfect doping concentration to maximize zT regardless of the scattering mechanism. This almost one order of magnitude increase in zT is due to large electrical conductivities from oxygen co-doping as well as low thermal conductivities from alloy scattering. The maximum power factor reached was $77.98 \times 10^{-4}$ $W/mK^2$ at 300K for $In_{0.32}Ga_{0.68}N$ alloys at a carrier concentration ~$6.25 \times 10^{20}$ $cm^{-3}$. This work indicates that $In_xGa_{1-x}N$ alloys have great potential for thermoelectric applications especially at a high temperature range.

Thermoelectric (TE) devices have the ability to directly convert heat into electricity using the Seebeck effect.[1–3] Solid-state devices such as thermoelectric generators (TEGs) have no moving parts, high reliability, and are potentially environmentally friendly depending on the materials used. Despite these attractive features, the applications of TE technology is limited by the use of toxic, rare earth, and expensive TE materials (e.g. $Bi_2Te_3$, SiGe).[4,5] Therefore, it is necessary to search for materials that can be used for TE applications beyond these toxic and rare earth elements.[6]

III-Nitride semiconductors have gained interest due to their promising TE properties and ability to tune their band gap for wide temperature ranges. Thus, their electronic properties can be tuned from an insulator behavior to a metallic behavior by manipulating their crystal structures chemical compositions, and doping concentrations.[1]

The efficiency of a TE material is determined by the figure of merit,

$$zT = \frac{S^2 \sigma T}{\kappa_L + \kappa_e} \qquad (1)$$

where S is the Seebeck coefficient, $\sigma$ is the electrical conductivity, $\kappa_L$ is the lattice thermal conductivity, $\kappa_e$ is the electronic thermal conductivity, and T is the absolute temperature.[7,8] A higher zT indicates more heat can be converted into electricity, which is desired for TE applications. To achieve a high zT in TE materials, high $S$ and $\sigma$, and low $\kappa_L$ and $\kappa_e$, are needed.

The past few decades have seen zT increasing from under a value of one to over two, however these materials, such as skutterudites[9] and clathrates,[10] were often difficult to incorporate into viable device structures.[11] More importantly, the past zT gains have been primarily driven by a reduction in the lattice thermal conductivity of materials through the use of nano-structuring[12–18] or new materials that have an inherently low thermal conductivity due to large effective masses of their elements.[19] These advances however have not been translated into commercially available devices[20], mostly due to the difficulty in engineering the electronic components of zT, whose tightly coupled parameters that make gains in the overall efficiency difficult to achieve.[21] For any given material, higher carrier concentrations increase the electrical conductivity, but also increase the electronic thermal conductivity and often decrease the Seebeck coefficient. Similarly, electrical conductivity and lattice thermal conductivity can be related due to phonon scattering mechanisms. For instance, lower lattice thermal conductivity can be achieved by increased phonon scattering, which generally reduces the electrical conductivity.[22]

III-Nitride semiconductor materials have been widely used in applications such as light-emitting diodes (LEDs)[23], solar cells[24], and high power devices.[25] Due to their high thermal stability, tunable wide bandgap, low thermal conductivity, and being cost-effective and environmentally friendly, $In_xGa_{1-x}N$ has been explored by this group and others as a high temperature TE material.[1,26–30] Previous studies showed promising high zTs for $In_xGa_{1-x}N$ alloys for a wide range of temperatures. For instance, Pantha *et al.* reported a zT value of 0.08 at 300K and 0.23 at 450K for $In_{0.36}Ga_{0.64}N$.[31] Furthermore, high temperature TE properties were studied by Nakamura's group, with a zT that reached 0.34 at 875K.[32] Although several groups have studied TE properties of $In_xGa_{1-x}N$ grown on sapphire substrates, the overall zT is still too low due to the trade-off between electrical conductivity and Seebeck coefficient. This could be attributed to low crystal quality of materials, such as phase separation in InGaN, due to the large lattice mismatch between InGaN and sapphire substrates. To this end, a novel structure of $In_xGa_{1-x}N$ grown on a GaN/ZnO substrate is investigated here with a focus on the TE properties in this work. $In_xGa_{1-x}N$ samples were produced with different thicknesses grown on a ZnO substrate and all show a small lattice mismatch with almost no phase separation, which drastically improves the electrical conductivity with little deterioration of the Seebeck coefficient. A maximum zT value of 0.86 was obtained at 300K for $In_{0.32}Ga_{0.68}N$ at a carrier concentration ~$1.5 \times 10^{20}$ cm$^{-3}$. Compared to previous studies[3], about a five times larger zT than that of SiGe, the current state of the art high temperature material, and about a one order of magnitude higher zT than $In_xGa_{1-x}N$ at room temperature is achieved in this work. A generalized "b-factor" approach for materials quality evaluation is also discussed here to find the optimized carrier concentrations[33]. This work shows that $In_xGa_{1-x}N$ alloys have a great potential for new thermoelectric devices especially those that will operate at high ambient temperatures.

## Results

The sample structure of $In_xGa_{1-x}N$ alloys on GaN/ZnO template is shown in **FIG. 1(a).** $In_xGa_{1-x}N$ layer thicknesses vary from 70-645 nm with different growth conditions. **FIG. 1(b)** shows a high resolution TEM image for one of the thin film samples. Clear lattice fringes of the $In_xGa_{1-x}N$ layer are observed, which indicates a high quality crystal structure. Since ZnO and GaN have a wurtzite structure with similar lattice constants (less than 1.8%), it shows a small lattice mismatch between the GaN buffer layer and ZnO substrate.

**FIG. 2** shows the HRXRD ω-2θ scan of the (002) reflection for $In_xGa_{1-x}N$ alloys. All samples show two, well separated peaks, in which a sharp peak (002) from the ZnO substrate and a broader peak at a lower angle from the $In_xGa_{1-x}N$ layer can be seen. The unique aspect of these samples is that the indium concentration is controlled at 32% for all samples, while the doping concentration, as well as sample thickness, was varied. In addition, the composition of the $In_xGa_{1-x}N$ was uniform and showed no significant phase separation, which is unusual for InGaN alloys with above 15% In content. This set of samples is based on a previous study of thermoelectric properties with different In content, which indicated an In content around 30% can maximize thermoelectric properties of $In_xGa_{1-x}N$ alloys.[28]

The dependence of mobility on carrier density is shown in **FIG. 3**. Carrier concentrations were measured as $1.39 \times 10^{18}$ cm$^{-3}$, $4.80 \times 10^{19}$ cm$^{-3}$, $1.50 \times 10^{20}$ cm$^{-3}$, and $6.25 \times 10^{20}$ cm$^{-3}$ for the four different samples shown in **TABLE 1**, while composition was held constant at around 32% indium. With increasing carrier density, the mobility increases up to 217 cm$^2$V$^{-1}$s$^{-1}$ and then decreases to 118 cm$^2$V$^{-1}$s$^{-1}$. The mobility increases due to the increased crystallinity of materials through the annealing process during growth. At very high carrier concentrations, the mobility decreases due to ionized impurity scattering becoming more dominant.

The electrical conductivity of $In_xGa_{1-x}N$ alloys as a function of carrier concentration is shown in **FIG. 3**. Although the mobility increases then decreases, the overall electrical conductivity in equation (2) and **FIG. 3** continues to increase, showing high electrical conductivity in $In_xGa_{1-x}N$ alloys using ZnO substrates. The conductivity depends on the cross-plane thickness of the sample. Since it is uncertain if parallel conduction through the GaN and ZnO layers is taking place, the measured electrical conductivity was scaled by the thickness of the entire sample. In this way, we have treated the entire structure as a single, composite material, which yields a conservative estimate to the electrical conductivity, power factor (PF), and zT.

The Seebeck coefficient and PF values of $In_xGa_{1-x}N$ alloys as a function of carrier concentration are plotted in **FIG.4**. With an increase of carrier concentration, the Seebeck coefficient decreases from -282 μV/K to -240 μV/K. This results in a peak PF value ~ 77.98×10⁻⁴ W/mK² at a carrier concentration of $6.25×10^{20}\,cm^{-3}$.

The carrier concentration dependence of the total thermal conductivity is plotted in **FIG.5**. The total thermal conductivity consists of contributions from both electron and phonon transport defined as[34],

$$\kappa_{tot} = \kappa_L + \kappa_e . \qquad (2)$$

Phonon scattering due to alloy disorder increases with the increase of carrier concentration[35]. Thus, the thermal conductivity shows a decreasing trend due to enhanced phonon scattering with the carrier concentration ranging from $1.39×10^{18}\,cm^{-3}$ to $1.50×10^{20}\,cm^{-3}$. It is interesting to note that when a carrier concentration of $6.25×10^{20}\,cm^{-3}$ is reached, the total thermal conductivity increases. This is due to the electronic thermal conductivity playing a more significant role in the total thermal conductivity at this carrier concentration. The trade-off between electrical and thermal properties leads to a peak zT value of 0.86 for a carrier concentration ~$1.50×10^{20}\,cm^{-3}$ at room temperature.

**FIG. 6** shows room temperature zT versus carrier concentration. Data for $In_xGa_{1-x}N$ from other studies are also included for comparison. The maximum room temperature zT is 0.86, which is one order of magnitude higher than those from other studies for $In_xGa_{1-x}N$ alloys in which zT is around 0.086[28]. This giant zT increase is attributed to an enhanced PF with low thermal conductivity at high carrier concentrations while also maintaining a large In content around 32%. Typically with In contents higher than 15%, phase separation will occur in the materials that reduces the crystal quality. However, using a ZnO substrate leads to a low lattice mismatch and no phase separation that enhances the TE properties of $In_xGa_{1-x}N$ alloys. More interestingly, a low thermal conductivity and high electrical conductivity are simultaneously reached without phase separation, which is not common as observed in other studies. The maximum room temperature zT value of these $In_{0.32}Ga_{0.68}N$ alloys is five times larger than SiGe alloys, which have a zT value of 0.16 at 300K[13].

**Discussion**

Once the material is synthesized and characterized by measuring the Seebeck coefficient, electrical, and thermal conductivities, two questions have been investigated to aid future materials design and fabrications, including what is the maximum zT achievable for this material at a given lattice thermal conductivity; and how should the doping/carrier concentration be adjusted to optimize zT? Due to the complexity of producing $In_xGa_{1-x}N$, the optimization of these TE materials tends to be empirical in nature. The technique utilized here is within the framework of a generalized b-factor approach[33,36]. Using the experimentally measured Seebeck, electrical and thermal conductivities, carrier concentrations, and assuming a power law scattering exponent, r, given by

$$\lambda(E) = \lambda_0 \left[ \left( E - E_C \right) \middle/ k_B T \right]^r \qquad (3)$$

with a "total" b-factor given as

$$b_{tot} \equiv \frac{\sigma T}{\kappa_e + \kappa_L}\left(k_B/q\right)^2$$

$$(4)$$

we arrive at an estimate for the doping necessary to maximize zT for each sample (all samples in theory "should" fall on the same curve, however experimental realities prohibit this). The results are shown in TABLE 2 for two different types of scattering, r = 0 (Acoustic Deformation Potential ADP), and r = 2 (Ionized Impurity). The theoretical doping concentrations needed to obtain a maximum zT value in TABLE 2 are very close to the values found from experiment for both types of scattering mechanisms. This shows all samples tested are close to the doping concentration that maximizes their zT, regardless of the scattering mechanism. It can be seen in **FIG. 7** that all the samples could benefit from a small increase in carrier concentration, however the increase in zT would be very small.

A large b-factor value is all that is important to increasing zT in the parabolic band model [37,38]. Once a semiconductor material is optimally doped, a large electrical conductivity and low thermal conductivity is all that is important to increasing zT, as can be seen from equation 4. The benefits gained to the electrical conductivity in this study from improved crystal quality of the samples outweighs the possible reduction of the b-factor due to an increase in the lattice thermal conductivity due to improved crystal quality. From the same procedure to estimate the theoretical doping, we also can estimate the lattice thermal conductivity from the experimentally obtained total thermal conductivity. The results are shown in TABLE 2. The largest zT of 0.86 at room temperature has the smallest estimated lattice thermal conductivity for either scattering assumption (i.e. r = 0 and r = 2).

**Conclusion**

In summary, the effect of carrier concentration on TE properties are comprehensively investigated in MOCVD-grown $In_xGa_{1-x}N$ alloys on GaN/ZnO substrates. One order of magnitude enhancement in the thermoelectric properties of $In_xGa_{1-x}N$ was obtained due to three main factors: 1) large electrical conductivities from a significant improvement in $In_xGa_{1-x}N$ material quality, i.e. the suppression of phase separation; 2) optimally doped carrier concentrations from oxygen and Si co-doping; and 3) relatively low thermal conductivities from alloy scattering. The high electrical conductivity was achieved without deteriorating the Seebeck coefficient. The optimum carrier concentration is found to be $\sim1.5\times10^{20}$ cm$^{-3}$ reaching a maximum zT value of 0.86 at room temperature, which is five times larger than SiGe (zT$\sim$0.16 at 300K). With higher ambient temperatures, zT is expected to increase due to a reduced lattice thermal conductivity as well as larger possible $\Delta$Ts. The experimental results indicate that $In_xGa_{1-x}N$ alloys grown on ZnO substrates have high potential for TE applications at room temperature, creating a new research direction in TE materials that has been conventionally focused in narrow bandgap materials. In addition, $In_xGa_{1-x}N$ alloys are widely used in lasers, optoelectronics, and solar cells so the device fabrication techniques are well developed, and it can be easily transferred to TE devices. As such, $In_xGa_{1-x}N$ alloys will provide a long term cost-effective solution at both the material and device level while offering several advantages over conventional TE materials, including high temperature operation and chemical stability, non toxicity, and earth abundancy.

**Methods**

**Metal-organic chemical vapor deposition (MOCVD) Growth.** Undoped and Si-doped $In_xGa_{1-x}N$ alloys were grown on a c-plane ZnO substrate on thin 30-50 nm undoped GaN templates using metal organic chemical vapor deposition (MOCVD). A wide range of growth conditions was explored for the different samples, including varying growth rates, dopant concentrations, and layer thicknesses. A theoretical work on calculated TE properties of $In_xGa_{1-x}N$ has been studied to elucidate the most promising composition and carrier densities using the Boltzmann transport model that considered different scattering mechanism.[39] The precursors for Ga, In, N, and Si were trimethylgallium (TMGa TEGa), trimethylindium (TMIn), ammonia ($NH_3$), and silane, respectively. Hydrogen and nitrogen were used as the carrier gases and silicon was used as n-type dopants with controlled carrier concentrations ranging from $10^{18}$-$10^{20}$ $cm^{-3}$. The indium concentration was determined from the peak value of the high-resolution X-ray diffraction (HRXRD) spectra in 2θ-ω scan mode and applying Vegard's law. The bandgap of $In_xGa_{1-x}N$ is around 3.25 eV measured by photoluminescence spectroscopy at room temperature. The thickness of the films was measured by spectral reflectometry. A high-resolution transmission electron microscopy was used to investigate crystal quality of the thin film samples.

**Thermoelectric properties characterizations.** The electrical conductivity, carrier density, and mobility were measured by using in-plane Van der Pauw Hall-effect measurements. In-plane Seebeck coefficients were obtained by standard temperature gradient methods.[3,28] Each sample was coated with a ~70nm thick layer of Al at room temperature to prepare the samples for thermal conductivity measurements. The cross-plane thermal conductivities were measured by time-domain thermo-reflectance (TDTR) from 300-500K with 5-10 MHz pump modulation frequency. TDTR[40,41] is an optical pump-probe technique to measure the nano-scale thermal conductivity of thin film samples[42–44]. The detected probe signal as a function of the transient surface temperature was fitted to the 3D thermal diffusion model based on the thermal quadrupoles[40,41]. The precision of the obtained thermal conductivity is approximately ±20%. Detailed measurement setup of TDTR is described in other references.[43,45,46] For zT calculations, the in-plane thermal conductivity is assumed to be equal to the cross-plane direction. In the absence of superlattices and other highly anisotropic structures, this is a commonly made assumption for thin film thermoelectrics[31,47] as well as in thermal conductivity studies of III-Nitride materials.[48–50] The structures under analysis have $In_xGa_{1-x}N$ layers larger than 70 nm, which can be considered larger than a superlattice layer or nanostructures and the assumption that the in-plane thermal conductivity is equal to the cross-plane is valid for our purposes here.

**Data Availability**

The datasets generated during and/or analyzed in the current study are available from the corresponding author on reasonable request.

**Acknowledgements.** The authors at Purdue University are grateful for the financial support from the National Science Foundation CAREER program (under Grants of CMMI − 1560834).

**Author contributions.** N. Lu. and I. F. contributed to the conception and design of experiments. Y. F., and B. K., conducted most of experiments, and electrical property characterization. E.W., and M. L. conducted b-factor modeling analysis. YR. K. and A. S. conducted thermal conductivity measurement. Y.F, and E. W. drafted the manuscript. N. L, and I. F. edited and revised the manuscript. N. L. supervised the project. All the authors discussed the results.

**Competing interests:** The authors declare no competing interests.

**Tables**

**TABLE 1.** Thermoelectric properties of $In_xGa_{1-x}N$ samples

| Seebeck Coefficient (µV/K) | Carrier Density (cm$^{-3}$) | Power Factor (10$^{-4}$ W/mK$^2$) | Mobility (cm$^2$V$^{-1}$s$^{-1}$) | Conductivity (Ω$^{-1}$cm$^{-1}$) | Thermal Conductivity (W/mK) | ZT | Sample thickness (nm) |
|---|---|---|---|---|---|---|---|
| -282 | 1.4E18 | 0.326 | 184 | 4.1 | 3.60±0.80 | 0.003 | 71 |
| -272 | 4.8E19 | 57.674 | 186 | 779.55 | 2.55±0.45 | 0.68 | 645 |
| -255 | 1.5E20 | 65.962 | 217 | 1014.41 | 2.30±0.45 | 0.86 | 130 |
| -240 | 6.25E20 | 77.98 | 118 | 1353.83 | 3.00±0.45 | 0.78 | 70 |

**TABLE 2.** Optimized carrier density and thermal conductivity based on "b-factor" simulations

| Sample | Carrier Density(cm$^{-3}$) | Optimized Carrier Density(cm$^{-3}$) r = 0 | Optimized Carrier Density(cm$^{-3}$) r = 2 | Extracted lattice thermal conductivity (W/mK) r = 0 | Extracted lattice thermal conductivity (W/mK) r = 2 |
|---|---|---|---|---|---|
| 2427gz | 4.8E19 | 8.70E19 | 9.30E19 | 2.19 | 1.89 |
| 2415gz | 1.5E20 | 1.95E20 | 2.05E20 | 1.83 | 1.45 |
| 2440gz | 6.25E20 | 7.37E20 | 8.16E20 | 2.36 | 1.88 |

**Figures:**

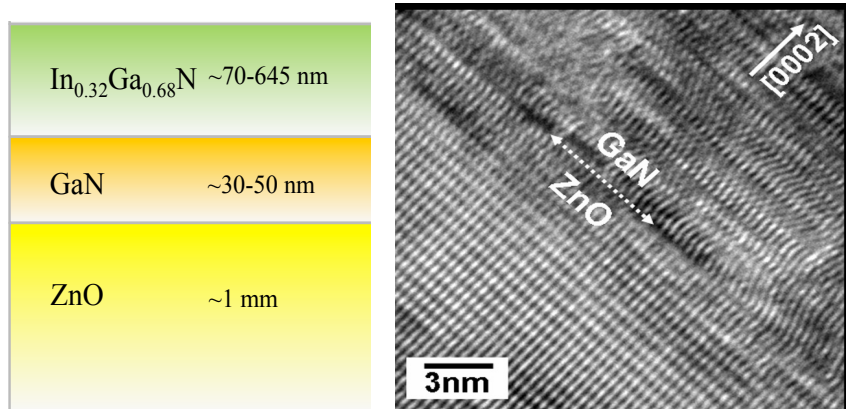

FIG. 1. Sample Structure of $In_xGa_{1-x}N$ thin films grown on ZnO substrate (1 mm) with GaN buffer layer (30-50nm) and a High Resolution TEM image of $In_xGa_{1-x}N$.

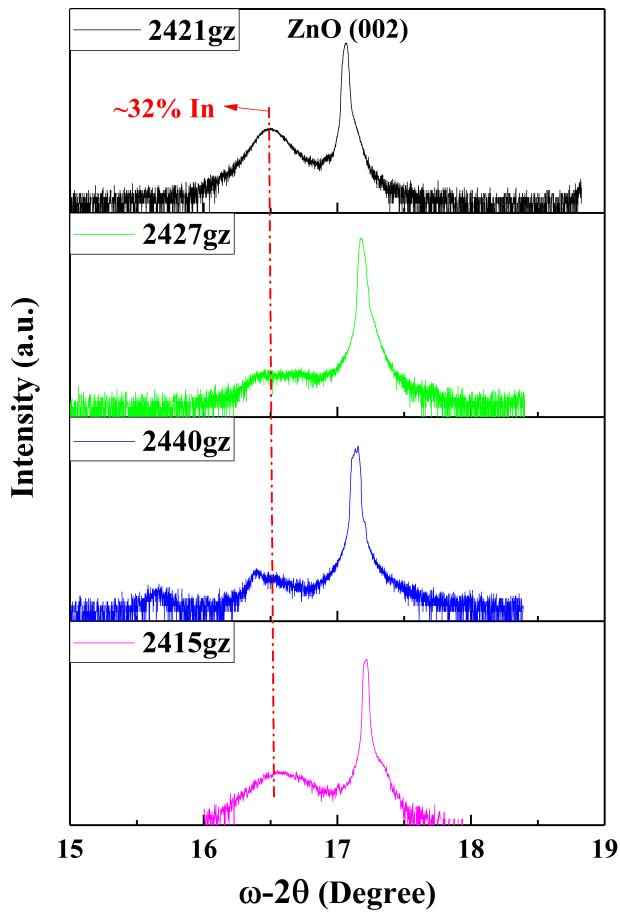

FIG.2. HRXRD 2θ-ω scan near (002) Bragg reflection plane for $In_xGa_{1-x}N$ thin films

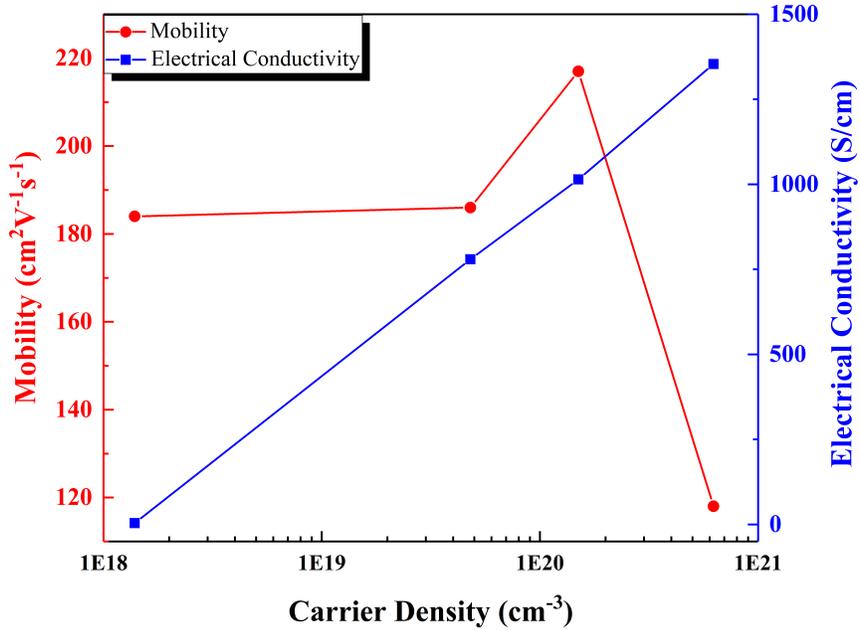

FIG.3. Mobility and electrical conductivity of In$_x$Ga$_{1-x}$N vs carrier density at room temperature with the thickness ranges from 70-645 nm

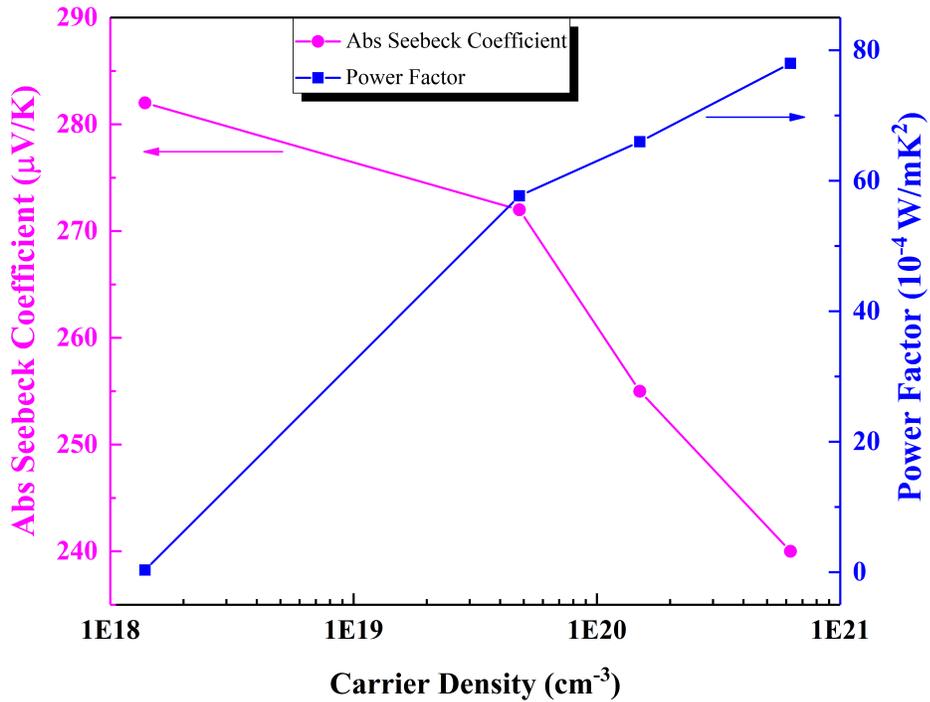

FIG. 4. The in-plane Seebeck coefficient measurement (pink line) and calculated power factor (blue line) of In$_x$Ga$_{1-x}$N vs carrier density at room temperature

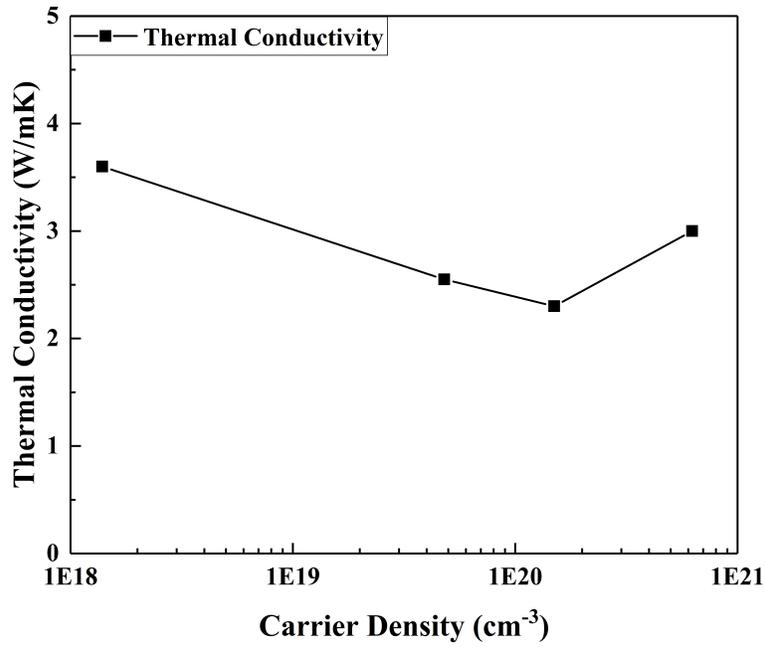

FIG. 5. Thermal conductivity of In$_x$Ga$_{1-x}$N vs carrier density at room temperature

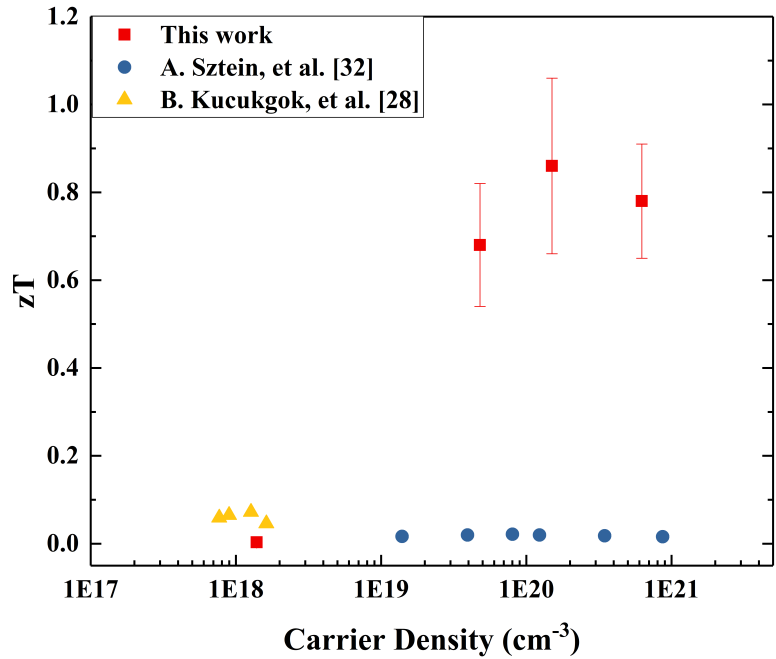

FIG. 6. Figure of Merit of In$_x$Ga$_{1-x}$N vs carrier density at room temperature, along with literature zT values for In$_x$Ga$_{1-x}$N

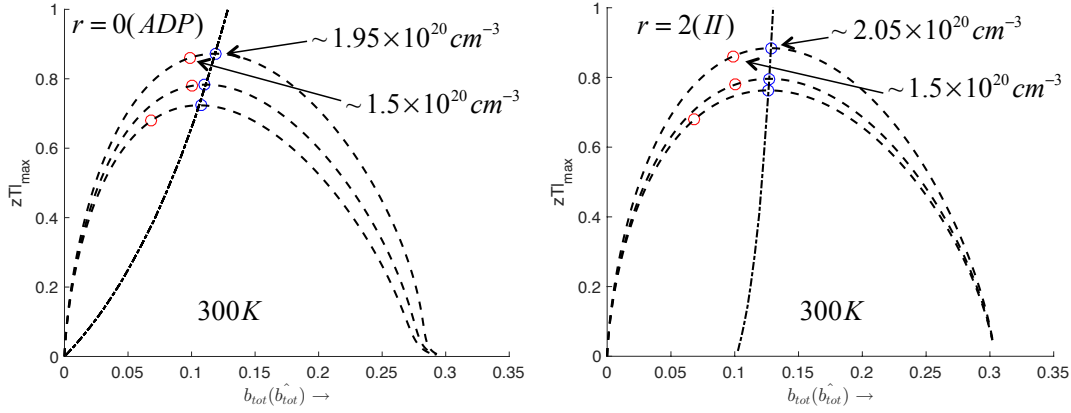

FIG. 7. Plot of the experimental zT values (red circles) versus the total b-factor for acoustic deformation potential scattering (left) and ionized impurity scattering (right) at 300K. The optimum zT curve is shown as a dotted-dashed line in both plots, with the optimum zT value for each sample shown with blue circles on the optimum zT curve. The dashed line for each sample corresponds to the zT value if only the Fermi level is varied for each sample.